\documentclass[aps,
  pra,                
  reprint
]{revtex4-2}
\usepackage[normalem]{ulem}
\usepackage[frak=esstix]{mathalpha} %lighter mathfrak

\usepackage{amsmath, amssymb}
\usepackage{graphicx}
\usepackage{bm}
\usepackage{mathrsfs}
\usepackage{lineno}
\usepackage{setspace}
\usepackage{comment}
\usepackage{scalerel}
\usepackage{xcolor}
\usepackage{slashed}
\usepackage{upgreek}

\begin{document}

\title{Arbitrary-Velocity Volkov Wavepackets}
\author{D. Ramsey}
\affiliation{Laboratory for Laser Energetics, University of Rochester, Rochester, New York 14623-1299, USA}
\author{J. McKeown}
\affiliation{Laboratory for Laser Energetics, University of Rochester, Rochester, New York 14623-1299, USA}
\author{J. P. Palastro}
\affiliation{Laboratory for Laser Energetics, University of Rochester, Rochester, New York 14623-1299, USA}

\date{\today}
%%%%%%%%%%%%%%%%%%%%%%%%%%%%%%%%%%%%%%%%%%%%%%%%%%%%%%%%%%%%%%%%
%% Commands
%%%%%%%%%%%%%%%%%%%%%%%%%%%%%%%%%%%%%%%%%%%%%%%%%%%%%%%%%%%%%%%%
\newcommand{\partialD}[1]{\ensuremath{\frac{\partial}{\partial #1}} }    
\newcommand{\partialDsq}[1]{\ensuremath{\frac{\partial^2}{\partial #1^2}} } 
\newcommand{\cpx}{\ensuremath{\tfrac{1}{2}i\kappa_0 w_0^2}}
\newcommand{\unitvec}[1]{\ensuremath{\bm{\hat{\mathrm{#1}}}}}
\newcommand{\bvec}[1]{\ensuremath{\bm{\mathrm{#1}}}}
\newcommand{\cycleavg}[1]{\left\llbracket #1 \right\rrbracket}

\newcommand{\toblue}[1]{{\color{blue}#1}}

\newcommand{\appropto}{\mathrel{\vcenter{
  \offinterlineskip\halign{\hfil$##$\cr
    \propto\cr\noalign{\kern2pt}\sim\cr\noalign{\kern-2pt}}}}}

%TC:ignore
\begin{abstract}
The evolution of a charged lepton in the field of an electromagnetic plane wave can be described as a superposition of Volkov states. Here we demonstrate that imposing specific momentum correlations among Volkov states produces a spatiotemporally structured wavepacket whose probability-density peak travels at an arbitrary, tailored velocity. This velocity can be chosen independently of both the field amplitude and the velocity expectation value. The imposed momentum correlations modify the expectation-value trajectory, providing a measurable signature of the arbitrary velocity within a physical observable.
\end{abstract}
%TC:endignore
\maketitle

\section{Introduction} 
The interaction of an electromagnetic field with a charged lepton induces spatiotemporal structure in the lepton's wavepacket. In the case of an electromagnetic plane wave, the wavepacket is expressible as a superposition of Volkov states \cite{Volkov1935, berestetskii1982quantum}. These states account for the field by modifying the bispinor and spatiotemporal phase of field-free solutions to the Dirac equation. The relative phases and amplitudes of the superposed states determine the initial structure of the wavepacket before it enters the field, while the Volkov modifications govern its ``dressed'' evolution within the field. This dressed evolution is the basis for the Furry picture of strong-field quantum electrodynamics \cite{furry1951bound}, where Volkov states are used in place of field-free solutions to calculate cross sections for processes such as nonlinear Compton scattering and Breit--Wheeler pair production  \cite{ReviewAntonino, gonoskov2022charged, fedotov2023advances}. 

Typically, the initial wavepacket is assumed to have a correlation among its momenta set only by the on-shell condition. However, the emergence of space--time structured light \cite{Piccardo2025, Abouraddy2025} and progress in wavefunction engineering \cite{okamoto2006quantum, mcmorran2011electron,lloyd2012quantized, shiloh2014sculpturing, kaminer2015self, harris2015structured, wong2021control, chirita2022transverse, campos2024nonspreading, velasco2025free} have motivated the exploration of wavepackets with additional correlations. In the context of light, an optical pulse with an appropriately chosen correlation in configuration or wavenumber space can exhibit an intensity peak that travels at an arbitrary velocity while maintaining a near-constant profile over long distances~\cite{sainte2017controlling, froula2018spatiotemporal,kondakci2019optical, palastro2020dephasingless,Caizergues2020, simpson2020nonlinear,jolly2020controlling, yessenov2022space,simpson2022spatiotemporal, ramsey2023exact, ASTRL, almeida2025universal, yessenov2025optical}. The extended interaction lengths and controllable velocities enabled by these pulses benefit a wide range of applications, including particle acceleration~\cite{palastro2020dephasingless,Caizergues2020,ramsey2020vacuum,palastro2021laser,liberman2025first}, radiation generation ~\cite{howard2019photon,ramsey2022nonlinear,kabacinski2023spatio,simpson2024spatiotemporal,fu2025steering,ramsey2025x}, detection of strong-field quantum electrodynamical phenomena~\cite{ formanek2024signatures,formanek2025enhanced}, and LIDAR~\cite{hall2025long}. 

Lepton wavepackets, like optical pulses, are solutions to a wave equation. As a result, the formalism underlying space--time structured light can be extended to matter waves \cite{palastro2024superluminal}. This extension was first used to demonstrate that correlations among the momenta of a field-free lepton wavepacket can produce a probability-density peak that travels at any velocity, including those exceeding the speed of light \cite{palastro2024superluminal}. It was later applied to a specific case in an electromagnetic plane wave, where a wavepacket with a negative momentum exhibited a peak traveling forward at the speed of light \cite{di2024radiation}. These realizations of space--time structured matter waves belong to the broader field of wavefunction engineering, which promises to enhance processes such as electron microscopy and electron holography \cite{okamoto2006quantum, mcmorran2011electron,lloyd2012quantized, shiloh2014sculpturing, kaminer2015self, harris2015structured, wong2021control, chirita2022transverse, campos2024nonspreading, velasco2025free}. 

Here, we show that the probability-density peak of a spatiotemporally structured lepton wavepacket can travel through an electromagnetic plane wave at any velocity, independent of its velocity expectation value. These wavepackets are constructed by prescribing specific correlations among the momenta of the constituent Volkov states. The correlations can be built into the initial wavepacket before it enters the electromagnetic field to produce a desired peak velocity within the field. The imposed correlations also modify the expectation-value trajectory, providing a measurable signature of the peak's velocity. As an extension, we show that conventional wavepackets, with no prescribed momentum correlations, acquire correlations between their dressed momenta upon interaction with an electromagnetic plane wave. 

The remainder of this article is organized as follows: Section \ref{sec:construct} describes the momentum-dependent amplitude required to generate a lepton wavepacket with a peak that travels at an arbitrary, prescribed velocity outside of a field. This description is then generalized to an arbitrary, prescribed velocity within the field. Section \ref{sec:dynamics} presents a calculation of the peak trajectory (Subsection A) and expectation trajectory (Subsection B). This is followed by specific examples that compare the two trajectories, including a description of how the imposed correlations modify the expectation-value trajectory (Subsection C). Section \ref{sec:conclude} discusses the natural emergence of space--time structure during the interaction of a charged lepton with an electromagnetic wave and concludes the manuscript.

Throughout, natural units ($\hbar = c = 1$) and the mostly-minus metric $(+,-,-,-)$ are used. Relativistic notation for superscripts and subscripts is not used. Four-vector products are denoted by $p\cdot x\equiv p^\mu x_\mu$, and an unindexed four-vector $p$ appearing alone denotes the four-vector $p^\mu$. Notational conventions are summarized in Table~\ref{tab:notation}. 

The considered interaction is between a charged, positive-energy lepton propagating along $x_3$ and an electromagnetic plane wave propagating in the positive $x_3$ direction. The plane wave is linearly polarized in the $x_1$ direction. A coordinate transformation $(x_0, x_3) \to (x_-, x_3)$ is employed, where $x_- \equiv x_0 - x_3$; any subsequent use of $x_0$ is understood to be implicitly defined through $x_-$ and $x_3$. Since expectation values evolve in $x_-$, instead of time, $x_0$, the bispinor inner product is constructed using $\gamma_- \equiv \gamma_0 - \gamma_3$  \cite{seipt2017volkov}. 

\begin{table}[ht]
\centering
\resizebox{\columnwidth}{!}{%
\begin{tabular}{ll}
\hline
\textbf{Symbol/Expression} & \textbf{Description} \\
\hline
$x = (x_0, \bvec{x})$, $\bvec{x} = (x_1, x_2, x_3)$ & Four-position \\
$p = (p_0, \bvec{p})$, $\bvec{p} = (p_1, p_2, p_3)$ & Four-momentum \\
$\bvec{a}_\perp = (a_1,a_2)$ & Transverse components of a three-vector\\
$k = \omega n$, $n = (1, 0, 0, 1)$ & Four-wavevector of the plane wave \\
$A = (0, A_1( x_-), 0, 0)$ & Electromagnetic four-potential \\
$\gamma=(\gamma_0,\gamma_1,\gamma_2,\gamma_3)$ & Dirac gamma matrices \\
$u^\dagger$ and $\bar{u} = u^\dagger \gamma_0$ & Hermitian conjugate and Dirac adjoint \\
$\bvec{a} \cdot \bvec{b} = a_1 b_1 + a_2 b_2 + a_3 b_3$ & Three-vector inner product \\
$a \cdot b = a_0 b_0 - \bvec{a}\cdot \bvec{b}$ & Four-vector inner product \\
$\tilde{a}$ & Field cycle-averaged quantity \\
\hline
\end{tabular}%
}
\caption{Summary of notation.}
\label{tab:notation}
\end{table}

\section{Wavepacket Construction}\label{sec:construct}
%NEED TO:
% *REF FIG 1 (ILLUSTRATION OF PROPAGATION) 
% *MAKE FIG 2 
\begin{figure}[th]
\centering
\includegraphics[width=0.5\textwidth]{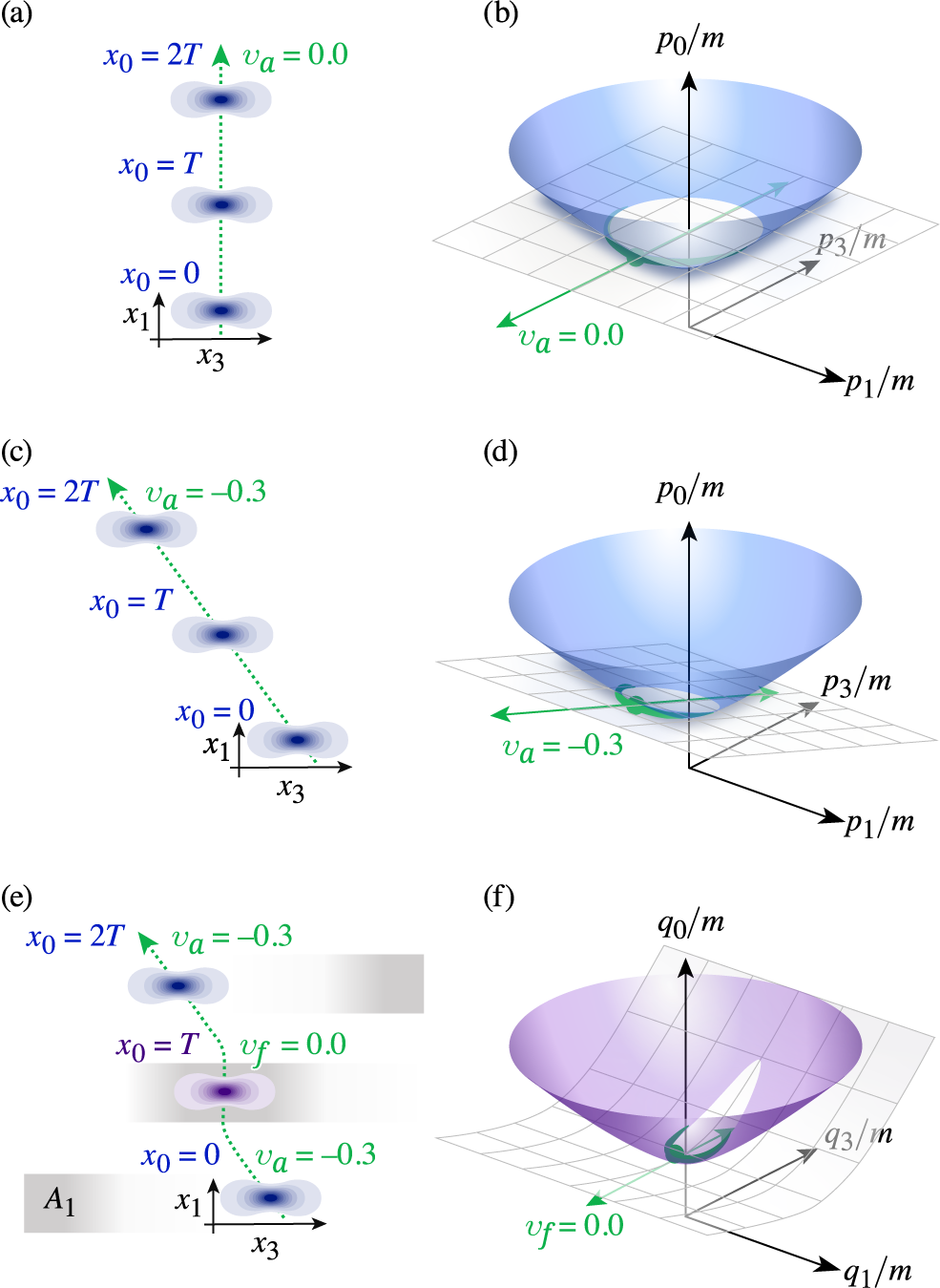}
\caption{Spatiotemporal evolution and momentum representations of partial wavepackets. (a,c,e) Snapshots at three times showing the advection of the amplitude $\bar{\Phi}\gamma_-\Phi$ for (a) a conventional wavepacket with $v_a = 0.0$; (c) an arbitrary-velocity wavepacket with $v_a = -0.3$; and (e) an arbitrary-velocity Volkov wavepacket with $v_a = -0.3$ outside the field and $v_{f*} = 0.0$ at the maximum vector potential $|e|\xi_*/m = 3.0$. The dotted green lines track the location of the peak amplitudes. The gray shading illustrates the location of the electromagnetic plane wave. (b,d) Corresponding momentum representations. The green curves indicate the intersection of the mass shell $E_{\bvec{p}}$ (blue) with the plane $\eta = p_0 - v_a p_3$ (translucent, with slope marked by the green arrow). (f) Dressed momentum representation of the Volkov wavepacket. Here the green curve indicates the intersection of the dressed mass shell $U_{\bvec{q}}$ (purple) and nonlinear surface $\lambda(\eta,q_-) = q_0 - v_{f*}q_3$ (translucent). For all panels, $p_-(\eta,0) = 3m$, yielding $\eta = 1.67m$ for $v_a = 0.0$ and $ {\eta} = 1.27m$ for $v_a = -0.3$.}%
\label{fig:1}
\end{figure}

The total wavepacket $\psi(x)$ of a charged lepton can be constructed from a superposition of ``partial'' wavepackets $\Phi(x,\eta)$ parameterized by $\eta$:
\begin{equation}\label{eq:Freewavepacket}
    \psi(x) 
    = \int  
     \mathcal{N}( \eta) \Phi(x,\eta) \mathrm{d} \eta,
\end{equation}
where $\mathcal{N}(\eta)$ is a scalar weighting function normalized such that $\int |\mathcal{N}(\eta)|^2  \mathrm{d}\eta = 1$ and $\int \bar{\psi}(x)\,\gamma_- \psi(x) \, \mathrm{d}^3\bvec{x} = 1$. In the absence of electromagnetic fields,
\begin{equation}\label{eq:Partialwavepacket_OOF}
    \Phi(x,\eta) 
    =\frac{1}{(2\pi)^{3/2}} \int 
    \frac{u(p)}{\sqrt{2 p_-}}\delta(p_0-E_{\bvec{p}})  f(p,\eta) 
    \mathrm{e}^{-i p\cdot x} \mathrm{d}^4 p,
\end{equation}
where $p_- \equiv p_0-p_3$, $E_{\bvec{p}} \equiv (m^2 + |\bvec{p}|^2)^{1/2}$, and the bispinor $u$ is normalized according to $\bar{u} \gamma_- u = 2p_-$. The complex, scalar distribution $f(p,\eta)$ determines the $\eta$-dependent amplitude and phase of each momentum state composing the wavepacket.

The presence of an electromagnetic plane wave modifies the bispinor and phase of the partial wavepackets: 
\begin{equation}\label{eq:Partialwavepacket_INF}
    \Phi(x,\eta) 
    =\frac{1}{(2\pi)^{3/2}} \int 
    \frac{\mathcal{V}}{\sqrt{2 p_-}}\delta(p_0-E_{\bvec{p}})  f(p,\eta) 
    \mathrm{e}^{i S} \mathrm{d}^4 p,
\end{equation}
where 
\begin{equation}
\begin{split}
    &\mathcal{V}(p,x_-)
    \equiv  \left[ \mathbb{I} - \frac{ \gamma_- \gamma_1\, eA_1(x_-)}{2 p_-} \right]u(p)
    \end{split}
\end{equation}
is the dressed bispinor, $\mathbb{I}$ denotes the identity matrix, and 
\begin{equation}\label{eq:FieldAction}
S(p,x) = -p\cdot x+ \frac{1}{p_-} \int_{0}^{x_-} \Big[  e p_1 A_1(x_-') - \tfrac{e^2}{2} A_1^2(x_-') \Big] \mathrm{d}x_-'
\end{equation}
is the classical action. The resulting total wavepacket is a superposition of Volkov states. While $A_1$ may be any function of $x_-$, the examples presented here adopt the specific form 
\begin{equation}\label{eq:A1form}
A_1(x_-) = \xi(x_-)\cos(\omega x_- + \phi),
\end{equation}
where the profile $\xi(x_-)$ varies slowly compared to the period $2\pi/\omega$ and $\phi$ is an arbitrary phase. The following subsections develop the framework for arbitrary-velocity Volkov wavepackets by first reviewing the construction of conventional and field-free arbitrary-velocity wavepackets.

\subsection{Field-Free Conventional Wavepackets}
Outside of an electromagnetic field, conventional wavepackets are synthesized as superpositions of mono-energetic partial wavepackets with $\eta = p_0$. That is, $\Phi(x,\eta)$ is chosen to be time-independent up to a global phase advance, such that
\begin{equation}\label{eq:CpartialModeTrans}
    \Phi(x,\eta)-\Phi (x_0+T ,\bvec{x}_\perp, x_3,\eta)\mathrm{e}^{i\eta T} = 0,
\end{equation}
where $T$ is an arbitrary time interval. Taking the Fourier transform of Eq.~\eqref{eq:CpartialModeTrans} and multiplying by $\bar{u}\gamma_-$ yields the constraint
\begin{equation}\label{eq:CSpecReq}
f(p,\eta)\left[1-\mathrm{e}^{-i(p_0-\eta)T}\right]=0,
\end{equation}
which can only be satisfied if $\eta = p_0$ or equivalently $f(p,\eta) \propto \delta( \eta - p_0)$. After imposing the on-shell condition to replace $p_0$ with $E_{\bvec{p}}$ in Eq.~\eqref{eq:Partialwavepacket_OOF}, the longitudinal momentum $p_3$ is expressed in terms of $\eta$ and $\bvec{p}_\perp$ to enforce $\eta = E_{\bvec{p}}$:
\begin{equation}\label{eq:Ccritp}
    p_3(\eta,\mathrm{p}_\perp) =   
    \,\pm\,  \sqrt{\eta ^2 - (m^2 + \mathrm{p}_\perp^2)},
\end{equation}
where $\mathrm{p}_\perp \equiv |\bvec{p}_\perp| $. Here, the negative sign is selected so that all plane waves composing the wavepacket travel in the negative $x_3$ direction. 

Choosing partial wavepackets that have the time-translation property described by Eq. \eqref{eq:CpartialModeTrans} does not imply that the total wavepacket $\psi(x)$ will have the same property. However, under the right conditions, $\psi(x)$ will approximately hold this property for a finite duration. The first condition is that $\mathcal{N}(\eta)$ is narrowly peaked about some characteristic value of $\eta$ (e.g., the average energy in the case of $\eta = p_0$). The second condition is that there are no correlations between $\bvec{p}_\perp$ and $\eta$: 
\begin{equation}\label{eq:NoCorr}
    \langle \eta \bvec{p}_\perp\rangle - \langle \eta\rangle \langle \bvec{p}_\perp \rangle = \bvec{0},
\end{equation}
where the angled brackets denote an expectation value. In order to satisfy the second condition, $f(p,\eta)$ must take the form
\begin{equation}\label{eq:CmodalSolution}
f(p,\eta) = \Big| \frac{\mathrm{d}\eta}{\mathrm{d}p_3}\Big|^{1/2}\mathcal{T}(\mathbf{p}_\perp) \delta(\eta - p_0).
\end{equation}
The factor $|\mathrm{d}\eta/\mathrm{d}p_3|^{1/2}$ ensures the probability density is separable in $\bvec{p}_\perp$ and $\eta$. The transverse momentum distribution $\mathcal{T}(\bvec{p}_\perp)$ is normalized so that $\int |\mathcal{T}(\bvec{p}_\perp)|^2 \mathrm{d}^2\bvec{p}_\perp = 1$, and is sufficiently localized in $\bvec{p}_\perp$ to exclude evanescent waves ($\mathrm{p}_\perp^2 \ll \eta^2 - m^2$), making it independent of $\eta$. For computational efficiency, all numerical examples employ a reduced transverse dimensionality, with $|\mathcal{T}(\bvec{p}_\perp)|^2 = (w/\sqrt{2\pi}) \mathrm{exp}(-w^2 p_1^2/2)\delta(p_2)$, where $w$ characterizes the transverse momentum spread.

Figure 1(a) illustrates the time-independent amplitude of a conventional partial wavepacket $\bar{\Phi} \gamma_- \Phi$. Due to the transverse momentum spread ($\propto w^{-1}$), the partial wavepacket ``focuses'' along $x_3$ to a minimum transverse width $w$ and then diffracts. Figure~\ref{fig:1}(b) provides a geometric interpretation of the corresponding momentum distribution [Eq.~\eqref{eq:CmodalSolution}], with only the negative branch of Eq.~\eqref{eq:Ccritp} selected (green curve). The momentum domain lies at the intersection of the mass shell $E_{\bvec{p}}$ (blue) and the horizontal plane $\eta = p_0$ (translucent). 

\subsection{Field-Free Arbitrary-Velocity Wavepackets}
Arbitrary-velocity lepton wavepackets are superpositions of propagation-invariant partial wavepackets with $\eta = p_0 - v_a p_3$. The partial wavepackets satisfy
\begin{equation}\label{eq:STpartialModeTrans}
    \Phi(x,\eta)-\Phi (x_0+T ,\bvec{x}_\perp, x_3+ v_aT,\eta)\mathrm{e}^{i\eta T} = 0
\end{equation}
for an arbitrary velocity $v_a$ and time interval $T$ \cite{palastro2024superluminal, almeida2025universal}. This property implies that the amplitude of the partial wavepackets remains invariant in time along characteristics $x_3 - v_a x_0 = \mathrm{const}$. Note that setting $v_a=0$ recovers the same translational property expressed by Eq.~\eqref{eq:CpartialModeTrans}. Thus a conventional partial wavepacket can be recovered by setting $v_a=0$. 

Taking the Fourier transform of Eq.~\eqref{eq:STpartialModeTrans} and multiplying by $\bar{u}\gamma_-$ yields the constraint
\begin{equation}\label{eq:STSpecReq}
f(p,\eta)\left[1-\mathrm{e}^{-i(p_0 - v_a p_3 - \eta)T}\right]=0,
\end{equation}
which can only be satisfied if $\eta = p_0 - v_a p_3$ or $f(p,\eta) \propto \delta(\eta - p_0 + v_a p_3)$. After imposing the on-shell condition to replace $p_0$ with $E_{\bvec{p}}$, $p_3$ is expressed in terms of $\eta$ and $\mathrm{p}_\perp$ to enforce $\eta =  E_{\bvec{p}} -v_a p_3$:
\begin{equation}\label{eq:STcritp}
    p_3(\eta,\mathrm{p}_\perp)
    = \eta  v_a \upgamma_a^2 
    \,\pm\,  \sqrt{(\eta \upgamma_a^2)^2 - \upgamma_a^{2}(m^2 + \mathrm{p}_\perp^2)},
\end{equation}
where $\upgamma_a \equiv (1-v_a^2)^{-1/2}$. Equation \eqref{eq:STcritp} shows that even though there are no correlations between $\eta$ and $\mathbf{p}_\perp$ [Eq.~\eqref{eq:NoCorr}], the on-shell condition and constraint $\eta =  E_{\bvec{p}} -v_a p_3$ impose a correlation between $p_3$ and $\mathrm{p}_\perp$. The sign is chosen based on the value of $v_ a$. For $|v_a|\leq1$, the negative branch ensures that the longitudinal momenta are negative in a Lorentz frame moving with velocity $v_a$. For $|v_a|>1$, the positive branch ensures that the longitudinal momenta are negative in a Lorentz frame moving with velocity $1/v_a$. These two cases correspond to wavepackets with time-independent and longitudinal-coordinate-independent ``foci" in the Lorentz frame, respectively (see Ref.~\cite{palastro2024superluminal} for details).

As before, the total wavepacket $\psi(x)$ will approximately hold the property described by Eq. \eqref{eq:STpartialModeTrans} if $\mathcal{N}(\eta)$ is narrowly peaked about a characteristic value of ${\eta}$ and $\langle \eta \bvec{p}_\perp\rangle - \langle \eta\rangle \langle \bvec{p}_\perp \rangle = \bvec{0}$. This latter condition requires that $f(p,\eta)$ take the form
\begin{equation}\label{eq:STmodalSolution}
f(p,\eta) = \Big| \frac{\mathrm{d}\eta}{\mathrm{d}p_3}\Big|^{1/2}\mathcal{T}(\mathbf{p}_\perp) \delta(\eta - p_0 + v_ap_3),
\end{equation}
which builds the correlation into the wavepacket. The transverse momentum distribution $\mathcal{T}(\bvec{p}_\perp)$ is again normalized so that $\int |\mathcal{T}(\bvec{p}_\perp)|^2\mathrm{d}^2\bvec{p}_\perp = 1$ and excludes evanescent waves ($\mathrm{p}_\perp^2 \ll \upgamma_a^{2}\eta^2 -m^2$).

Figure 1(c) illustrates the time-dependent amplitude of an arbitrary-velocity partial wavepacket  $\bar{\Phi} \gamma_- \Phi$ with a peak amplitude that moves at $v_a = -0.3$. Like the conventional case, the space--time wavepacket has a minimum width $w$ determined by its transverse momentum spread. However, now the ``focusing" and diffraction occur with respect to the  coordinate $x_3-v_a x_0$. Figure~\ref{fig:1}(d) provides the geometric interpretation of the momentum distribution. The domain lies along the negative branch of Eq.~\eqref{eq:STcritp} (green curve), corresponding to the intersection of the mass shell $E_\mathbf{p}$ (blue) and the plane $\eta = p_0 - v_a p_3$ (translucent). Along the intersection
\begin{equation}\label{eq:STderivative}
    \left(\frac{\partial {E}_\mathbf{p}}{\partial{{p}}_3}\right)_\eta =  v_a,
\end{equation}
where $()_{\eta}$ denotes that the derivative is taken with $\eta$ held fixed. This relation indicates that $f(p,\eta)$ features a unique correlation among the momenta, i.e., $\partial {E}_\mathbf{p}/\partial{{p}}_3|_\eta \neq 0$. Nevertheless, when superposing partial wavepackets, $\eta$ varies on the mass shell, such that $\partial E_{\bvec{p}}/\partial p_3 = p_3/E_{\bvec{p}}$. Thus, while $v_a$ can take any value, the expectation value of the longitudinal velocity, $\langle p_3/E_{\bvec{p}} \rangle$, is always subluminal.

\subsection{Arbitrary-Velocity Volkov Wavepackets}
In the presence of an electromagnetic plane wave, field-dependent terms in the phase $S$ [Eq.\eqref{eq:FieldAction}] modify the constraint required for propagation invariance. To determine this constraint, consider a charged lepton interacting with a four-potential that has a constant strength $\xi = \xi_*$, such that $A_1(x_-) = \xi_* \cos(\omega x_- +\phi)$. The partial wavepackets are chosen to be propagation-invariant up to a global phase advance when averaged over a cycle of the potential:
\begin{equation}\label{eq:LepInvarCon}
\int_0^{\frac{2\pi}{\omega}} \left[ \Phi(x,\eta) - \Phi(x_0+T, \bvec{x}_\perp, x_3+v_{f} T,\eta) e^{i\eta T} \right] \mathrm{d}x_- = 0,
\end{equation}
where $T$ is an arbitrary time interval and $v_f$ is the arbitrary velocity in the field. Taking the Fourier transform of Eq.~\eqref{eq:LepInvarCon}, multiplying by $\bar{u}\gamma_-$, and applying the Jacobi–Anger expansion to simplify the cycle average yields the constraint 
\begin{equation}\label{eq:InFieldinvarReq}
f(p,\eta) \left[ 1 - \mathrm{e}^{-i( q_0 -v_{f} q_3 - \eta) T} \right] = 0.
\end{equation}
Here, the field-dressed four-momenta $q$ are defined as
\begin{equation}\label{eq:fielddressedmo}
q \equiv p + \frac{e^2 \xi_*^2}{4 p_-} n,
\end{equation}
subject to the dressed-shell condition $q_0 = U_{\bvec{q}}\equiv ( 
 m^2 + e^2\xi_*^2/2 + |\bvec{q}|^2 )^{1/2}$ \cite{berestetskii1982quantum}. The constraint expressed by 
Eq.~\eqref{eq:InFieldinvarReq} can only be satisfied if $f(p,\eta) \propto \delta(\eta - q_0 + v_f q_3)$, which shows that propagation invariance in an electromagnetic plane wave requires correlations among the dressed momenta.

At this stage, the wavepacket construction could proceed along the same path as in the previous subsections, with the momenta replaced by the dressed momenta. However, this approach has two practical limitations. First, a physically realizable wavepacket would typically be prepared outside the field, where imposing correlations among dressed momenta---whose dependence on the momenta is nonlinear---would be challenging. Second, the strength of a physical four-potential is never constant, i.e., $\xi$ depends on $x_-$, such that $A_1(x_-) = \xi(x_-)  \cos(\omega x_- +\phi)$. As a result, the propagation invariant property described by Eq. \eqref{eq:LepInvarCon} can only occur at a target potential strength $\xi = \xi_*$. 

Notably, these limitations do not apply in the special case of $v_f=1$, where $\eta =p_-$  and the probability-density peak travels at the constant velocity $v_a ={v_{f}}=1$ regardless of the local value of $\xi(x_-)$ \cite{di2024radiation}. This occurs because $p_-=q_-$ is a conserved quantity, ensuring that Eq. \eqref{eq:LepInvarCon} is satisfied across all potential strengths.

Here, the goal is to prepare a wavepacket with a linear momentum correlation outside the field, so its probability-density peak travels at a desired velocity $v_{f*}$ at a target strength $\xi_*$ inside the field. This is achieved by appropriately setting the out-of-field velocity $v_a$. To determine this velocity, the out-of-field plane $\eta = p_0 - v_a p_3$ is expressed in terms of the dressed momenta at $\xi = \xi_*$, yielding the surface
\begin{equation}\label{eq:NotAPlane}
    \lambda(\eta,q_-) - q_0 + v_{f*}q_3= 0,
\end{equation}
where
\begin{equation}
    \lambda  \equiv \left(\frac{1 -v_{f_*}}{1-v_a}\right) \eta - \left(\frac{v_a - v_{f_*}}{1-v_a} \right) q_-+ (1-v_{f*})\frac{e^2\xi_*^2}{4q_-}.
\end{equation}
The surface is locally planar with slope $v_{f*}$ relative to $q_3$ if
\begin{equation} \label{eq:deriv_va}
   \frac{\partial \lambda}{\partial q_3} =   \left(\frac{v_a - v_{f_*}}{1-v_a}\right) + (1-v_{f*})\frac{e^2\xi_*^2}{4q_-^2} = 0.
\end{equation}
When evaluated on-shell, $q_- = p_- = E_{\bvec{p}} - p_3(\eta, \mathrm{p}_\perp)$ [see Eq.~\eqref{eq:STcritp}], this condition depends on $\mathrm{p}_\perp$ and $\eta$. To first approximation, this dependence can be removed by evaluating $q_-$ at the expectation values $\langle \bvec{p}_\perp \rangle$ and $\langle \eta \rangle$, providing the out-of-field velocity 
\begin{equation}\label{eq:mapping}
    v_a \approx \frac{{v}_{f*} - (1-{v}_{f*})\tfrac{e^2\xi_*^2}{4 {\mathcal{P}}_-^2}}{1- (1-{v}_{f*})\tfrac{e^2\xi_*^2}{4 {\mathcal{P}}_-^2}},
\end{equation}
where ${\mathcal{P}}_- \equiv p_-(\eta = \langle \eta \rangle,\, \mathrm{p}_\perp = \langle \bvec{p}_\perp \rangle)$. With $v_a$ determined, the out-of-field distribution $f(p,\eta)$ is given by Eq.~\eqref{eq:STmodalSolution}. The resulting in-field distribution produces the desired linear correlation between the dressed momenta at $\xi = \xi_*$: $f(p,\eta) \appropto \delta[\Lambda - q_0 + v_{f*} q_3]$ in the vicinity of $q_-=\mathcal{P}_-$ and $\bvec{p}_\perp = \langle{\bvec{p}_\perp\rangle}$, where $\Lambda \equiv \lambda(\langle\eta\rangle, \mathcal{P}_-)$. Together with the on-shell condition, $q_0 = U_{\bvec{q}}$, the constraint imposed by the delta function reveals a correlation between $q_3$ and $\mathrm{p}_\perp$:
\begin{equation}\label{eq:STcritq}
\begin{split}
    q_3(\Lambda,\mathrm{p}_\perp)  
    \approx &\Lambda  v_{f*} \upgamma_{f*}^2 \\
    &\pm  \sqrt{(\Lambda \upgamma_{f*}^2 )^2 - \upgamma_{f*}^{2}(m^2 +  \tfrac{e^2}{2}\xi_*^2+ \mathrm{p}_\perp^2)},
\end{split}
\end{equation}
where $\upgamma_{f*} \equiv (1-v_{f*}^{2})^{-1/2}$, which is a dressed version of Eq.~\eqref{eq:STcritp}.

Figure~\ref{fig:1}(e) illustrates the dynamic amplitude of an arbitrary-velocity-Volkov partial wavepacket $\bar{\Phi} \gamma_- \Phi$. The wavepacket starts outside the field, where its peak amplitude travels at ${v_a=-0.3}$. Upon entering the field and encountering the target potential strength $|e|\xi_*/m = 3$, the peak travels at the designed velocity $v_{f*} = 0.0$. After leaving the field, the peak again travels at ${v_a=-0.3}$. Figures \ref{fig:1}(d) and \ref{fig:1}(f) provide the geometric interpretation. Outside the field, the momenta lie along the intersection of the mass shell and plane $\eta = p_0 - v_ap_3$ [Fig. \ref{fig:1}(d)]. Inside the field, the dressed momenta lie along the intersection of the dressed-shell and nonlinear surface $\lambda(\eta,q_-) = q_0 - v_{f*}q_3$ [Fig. \ref{fig:1}(f)]. The out-of-field velocity $v_a$ is chosen so that the nonlinear surface has the desired slope $v_{f*}$ in the vicinity of the momentum expectation values. That is, along the intersection and in the vicinity of $\bvec{p}_\perp = \langle \bvec{p}_\perp\rangle$,
\begin{equation}\label{eq:IdealCon}
   \left(\frac{\partial {U_{\bvec{q}}(\xi_*)}}{\partial{{q}}_3(\xi_*) }\right)_{\eta = \langle{\eta}\rangle} \approx  v_{f*}
\end{equation}
[see green arrow in Fig.~\ref{fig:1}(f)], which indicates that $f(p,\eta)$ locally produces the desired linear correlation among the dressed momenta.

\begin{figure}[t!]
\centering
\includegraphics[width=0.35\textwidth]{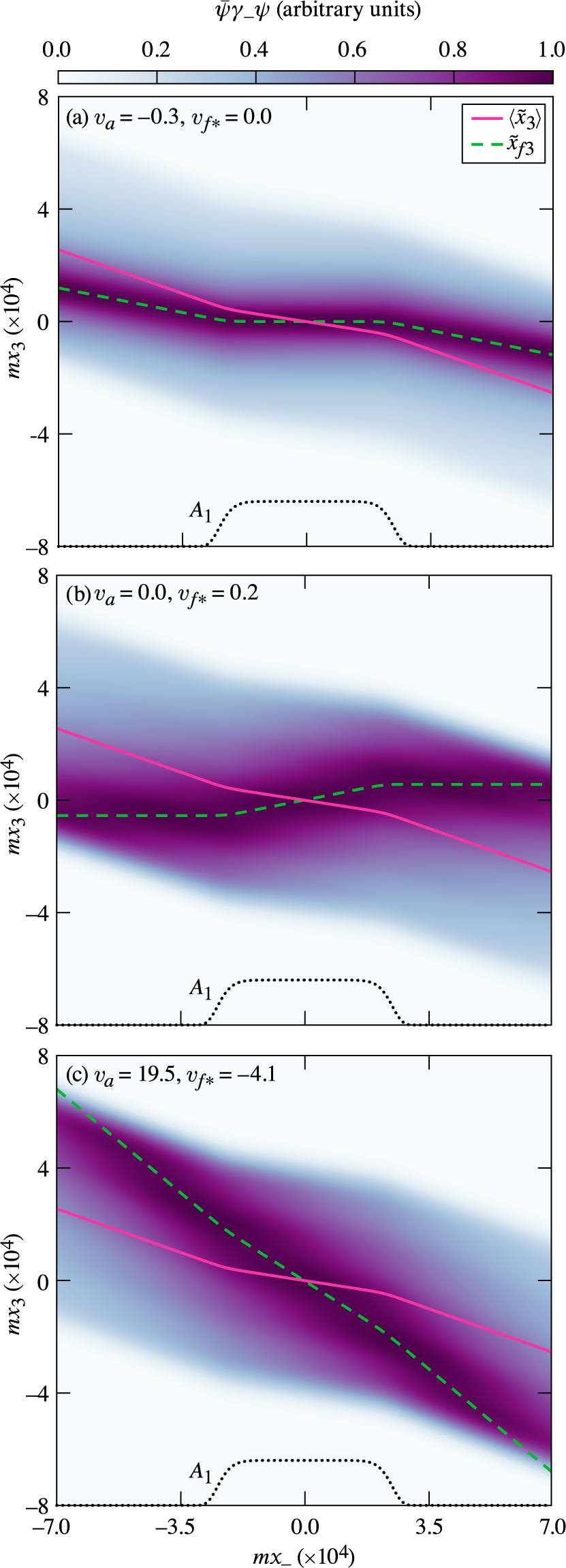}
\caption{Propagation of arbitrary-velocity Volkov wavepackets in $x_-$ and $x_3$. In panels (a), (b), and (c), the respective out-of-field velocities $v_a = -0.3,\, 0,$ and $19.5$ are chosen to yield the desired in-field velocities $v_{f*} = 0,\, 0.2,$ and $-4.1$ at the maximum potential strength $|e|\xi_*/m = 3$. For each case, the momentum content is peaked at ${\mathcal{P}}_- = 3m$ and $\mathrm{p}_\perp= 0$. The dotted black line shows the location and amplitude of the vector potential. The dashed green line marks the analytic, cycle-averaged longitudinal location of the probability-density peak $\tilde{x}_{f3}$, and the solid pink line the cycle-averaged expectation value ${\langle\tilde{x}_3\rangle}$. In all cases, the peak trajectory moves independently of the expectation trajectory. At $\xi = \xi_*$ and $\xi = 0$, the cycle-averaged expectation velocities are $\langle \tilde{v}_3 \rangle\approx - 0.24$ and $\langle \tilde{v}_3 \rangle\approx -0.8$, respectively.}  \label{fig:2}
\end{figure}

\section{Trajectories}\label{sec:dynamics}
This section examines the distinct dynamics of the expectation and peak-probability trajectories by presenting their equations of motion, comparing their evolution, and analyzing the influence of momentum correlations and $v_a$ on the expectation trajectory.

\subsection{Peak-Probability Trajectory}
When the weighting function $\mathcal{N}(\eta)$ is narrowly peaked about the expectation value $\langle {\eta}\rangle$, the phase $S$ [Eq.~\eqref{eq:FieldAction}] can be expanded about the intersection of $\langle{\eta}\rangle$ and the mass shell. When the paraxial approximation is valid, $\mathrm{p}_\perp^2  \ll (\langle \eta\rangle^2 \upgamma_a^2 - m^2)$, the phase can also be expanded to second order about $\mathrm{p}_\perp = \langle \mathrm{p}_\perp \rangle = 0$. Together, these approximations yield
\begin{equation}\label{eq:parxExpandS}
\begin{split}
    S \approx &\,{\mathcal{P}}_3 x_3 - {\mathcal{E}}x_0-\frac{1}{2\mathcal{P}_-}\mathcal{I}_2(x_-)\\
    +&p_1 x_1 + \frac{ p_1}{{\mathcal{P}}_-}\mathcal{I}_1(x_-) + p_2 x_2\\
    +&\frac{\mathrm{p}_\perp^2}{2(\mathcal{E}v_a-{\mathcal{P}}_3)}\left(x_3 -v_a x_0 -\frac{1-v_a}{2{\mathcal{P}}_-^2}\mathcal{I}_2(x_-)\right),
\end{split}
\end{equation}
where ${\mathcal{P}}_3 \equiv p_3(\eta=\langle\eta\rangle, \mathrm{p}_\perp=0)$ with $p_3$ given by Eq.~\eqref{eq:STcritp}, ${\mathcal{E}} \equiv (m^2 + \mathcal{P}_3^2)^{1/2}$, and
\begin{equation*}
\begin{split}
        \mathcal{I}_1(x_-)&=\int_0^{x_-} e A_1(x_-')\;\mathrm{d}x_-',\\
        \mathcal{I}_2(x_-)&=\int_0^{x_-} e^2  A_1^2(x_-')\;\mathrm{d}x_-'.
\end{split}
\end{equation*}
Although the final term has an apparent singularity at $v_a = {\mathcal{P}}_3/{\mathcal{E}}$, this value of $v_a$ lies outside the regime of validity of the approximations made to derive Eq.~\eqref{eq:parxExpandS} (see Appendix A).

Each term in Eq.~\eqref{eq:parxExpandS} can be understood by conceptualizing the wavepacket as an envelope modulated by a carrier wave. The carrier describes the motion of the phase fronts, which is governed by the terms of $\mathcal{O}(\mathrm{p}_\perp^0)$ [first line of Eq.~\eqref{eq:parxExpandS}]. The envelope describes the trajectory of the probability-density peak $\bvec{x}_{f}$, which is governed by the remaining terms [second and third lines  of Eq.~\eqref{eq:parxExpandS}]. The transverse and longitudinal locations of the peak are determined by the vanishing of the $\mathcal{O}(\bvec{p}_\perp)$ and $\mathcal{O}(\mathrm{p}_\perp^2)$ terms, respectively, for all $p$. For $v_a \neq 1$, this results in the equations of motion:
\begin{align}
    \frac{\mathrm{d}x_{f1}}{\mathrm{d}x_-} 
        &= -\frac{{eA_1}}{{\mathcal{P}}_-}, 
        \label{eq:xf} \\[6pt]
    \frac{\mathrm{d}x_{f2}}{\mathrm{d}x_-}  
        &= 0, 
        \label{eq:yf} \\[6pt]
    \frac{\mathrm{d}x_{f3}}{\mathrm{d}x_-}  
        &= \frac{{v_{a}} }{1-{v_a}} 
           + \frac{e^2 A_1^2}{2{\mathcal{P}}_-^2}.
        \label{eq:zf}
\end{align}
For $v_a=v_f=1$, $\mathrm{d}x_{f3}/\mathrm{d}x_0 = 1$  
and the longitudinal trajectory is non-oscillatory at any field amplitude \cite{di2024radiation}. With $A_1 = \xi(x_-)\cos(\omega x_- + \phi)$, the cycle average (denoted by a tilde) of Eq.~\eqref{eq:zf} becomes
\begin{equation}\label{eq:lightframefocal}
        \frac{\mathrm{d}\tilde{x}_{f3}}{\mathrm{d}x_-}  
        = \frac{{v_{f}}(x_-) }{1-{v_f}(x_-)}, 
\end{equation}
where
\begin{equation}\label{eq:vf_general}
    v_f(x_-) = \frac{v_a + (1-v_a)\tfrac{e^2}{4 {\mathcal{P}}_-^2}\xi^2(x_-)}{1+ (1-v_a)\tfrac{e^2}{4 {\mathcal{P}}_-^2}\xi^2(x_-)}.
\end{equation}
Equation~\eqref{eq:vf_general} shows that the longitudinal velocity of the probability-density peak depends on the local value of $\xi$ and confirms that a constant velocity $v_{f*}$ can only be achieved at a target value $\xi = \xi_*$. Note that even when $v_a = 0$, the probability-density peak follows a trajectory determined by the local potential strength.

\subsection{Expectation-Value Trajectory}
The equations of motion for the position expectation values follow from
\begin{equation*}
       \frac{\mathrm{d}\langle \bvec{x}\rangle}{\mathrm{d}x_-}
 = \frac{\mathrm{d}}{\mathrm{d} x_-} \int \bvec{x} \;\bar{\psi}(x)\, \gamma_- \, \psi(x)\, \mathrm{d}^3\bvec{x}. 
\end{equation*}
Componentwise, this gives
\begin{align}
     \frac{\mathrm{d}\langle x_1\rangle}{\mathrm{d}x_-}  &=-\int  \left(\frac{eA_1}{{p}_-}\right) 
       |\mathcal{N}(\eta)|^2 |\mathcal{T}(\bvec{p}_\perp)|^2 \,\mathrm{d}^2\bvec{p}_\perp \, \mathrm{d}\eta,  \label{eq:EXPx} \\[6pt]
    \frac{\mathrm{d}\langle x_2\rangle}{\mathrm{d}x_-} &=0, \label{eq:EXPy} \\[6pt]
   \frac{\mathrm{d}\langle x_3\rangle}{\mathrm{d}x_-} &= \int \left(\frac{p_3}{p_-} +\frac{e^2A_1^2 }{2p_-^2}\right)
       |\mathcal{N}(\eta)|^2 |\mathcal{T}(\bvec{p}_\perp)|^2 \, \mathrm{d}^2\bvec{p}_\perp \, \mathrm{d}\eta \label{eq:EXPz}, 
\end{align}
where $\mathcal{T}(\bvec{p}_\perp)$ is assumed to be an even function of $\bvec{p}_\perp$ and the momenta are on-shell: $p_-(\eta,\bvec{p}_\perp) = E_{\bvec{p}} - p_3(\eta,\bvec{p}_\perp)$ with $p_3(\eta,\bvec{p}_\perp)$ given by Eq.~\eqref{eq:STcritp}. 

An approximate expression for the cycle-averaged longitudinal velocity is obtained by expanding the terms in parentheses about $\eta = \langle \eta \rangle$ and $\mathrm{p}_\perp = 0$ to second order, leading to
%checked EQ (3/20) [X]
\begin{equation}\label{eq:lighfvel}
\begin{split}
    {\frac{\mathrm{d}\langle \tilde{x}_3\rangle}{\mathrm{d}x_-}}
     \approx & \frac{v_{1\mathrm{D}}}{1-v_{1\mathrm{D}}}\\
       &+ \frac{1 - v_{f} v_{1\mathrm{D}}}{2{\mathcal{P}}_-^2  (v_{f} - v_{1\mathrm{D}})}\int \mathrm{p}_\perp^2 |\mathcal{T}(\bvec{{p}}_\perp)|^2 \, \mathrm{d}^2\bvec{p}_\perp,
       \end{split}
\end{equation}
where
\begin{equation}\label{eq:v1D}
    v_{\mathrm{1D}} = \frac{\mathcal{P}_3/\mathcal{E} + (1-\mathcal{P}_3/\mathcal{E} )\tfrac{e^2}{4 {\mathcal{P}}_-^2}\xi^2(x_-)}{1+ (1-\mathcal{P}_3/\mathcal{E} )\tfrac{e^2}{4 {\mathcal{P}}_-^2}\xi^2(x_-)}
\end{equation}
is the one-dimensional drift velocity of a planar wavepacket. The second term in Eq.~\eqref{eq:lighfvel} accounts for the effect of transverse momentum and depends explicitly on $v_{f}$ (and thus on $v_a$).  This dependence is a consequence of enforcing separability between $\eta$ and $\bvec{p}_\perp$, which necessarily introduces correlations among other momentum combinations. For example, a conventional wavepacket ($v_a = 0$) exhibits no correlation between $p_0$ and $\bvec{p}_\perp$, i.e., $\langle p_0 \bvec{p}_\perp\rangle - \langle p_0 \rangle \langle \bvec{p}_\perp \rangle = \bvec{0}$, whereas an arbitrary-velocity wavepacket ($v_a \neq 0$) does: $\langle p_0 \bvec{p}_\perp\rangle - \langle p_0 \rangle \langle \bvec{p}_\perp \rangle \neq \bvec{0}$. Note that the apparent singularity at $v_{1\mathrm{D}} = v_{f}$ is equivalent to the singularity observed in Eq.~\eqref{eq:parxExpandS} [cf. Eqs. \eqref{eq:lighfvel} and \eqref{eq:vf_general} and see Appendix A]. 
 
\subsection{Trajectory Comparison}

Figure~\ref{fig:2} displays the total probability density $\bar{\psi}\gamma_-\psi$ generated from Eq.~\eqref{eq:Freewavepacket} for three peak trajectories at $x_1 = x_{f1}(x_-)$. In panels (a)--(c), the out-of-field velocities $v_a = -0.3$, $0$, and $19.5$ are chosen to yield the in-field velocities $v_{f*} = 0$, $0.2$, and $-4.1$ at $\xi = \xi_*$, respectively. The dashed green line shows $\tilde{x}_{f3}$ and the solid pink line $\langle \tilde{x}_3 \rangle$. To emphasize that $\tilde{x}_{f3}$ is independent of $\langle \tilde{x}_3 \rangle$, the $\langle \tilde{x}_3 \rangle$ trajectories are chosen to be nearly identical in all three cases. In each case, $\mathcal{P}_- = 3m$, $w = 170m^{-1}$ and $\mathcal{N}(\eta)$ is defined so that its Fourier transform exhibits a region of near-constant amplitude in time, modeled here as a tenth-order super-Gaussian. An illustrative, short-duration vector potential with $\omega = 0.01 m$ is used for simultaneous visualization of the in-field and out-of-field dynamics within a single frame. The profile of the potential $\xi(x_-)$, indicated by the dotted black line, has a full width at half maximum of $4.8\times10^4 m^{-1}$ and a maximum strength $|e|\xi_*/m = 3$. 

Figure~\ref{fig:2}(a) shows the dynamics illustrated in Fig.~\ref{fig:1}(e), where a desired in-field velocity $v_{f*}=0.0$ is achieved by setting $v_a=-0.3$. Figure 2(b) demonstrates that a conventional wavepacket ($v_a = 0$) acquires a non-zero velocity ($v_f \neq 0$) upon entering the field. This behavior confirms the prediction of Eq.~\eqref{eq:vf_general} and establishes that wavepackets initialized without prescribed momentum correlations acquire correlations between their dressed momenta upon interaction with a field.

In each panel of Fig.~\ref{fig:2}, the peak trajectory is only distinguishable within an envelope that follows the expectation-value trajectory. The finite lifetime of the peak is due to interference among partial wavepackets, which restricts the translational property described by Eqs.~\eqref{eq:STpartialModeTrans} and \eqref{eq:LepInvarCon} to a finite duration. In all cases, the envelopes have equal lengths in $x_3$, and the peaks are apparent over the interval required to traverse this length. This is most clearly illustrated in Fig.~\ref{fig:2}(c), where the large disparity between the expectation velocity ($\langle \tilde{v}_{3} \rangle = -0.8$ outside the field) and the peak-probability velocity ($v_{f*} = -4.1$) results in a shorter lifetime compared to Figs.~\ref{fig:2}(a) and (b). A calculation of the lifetime is presented in Appendix B.

\begin{figure}[ht]
\centering
\includegraphics[width=0.45\textwidth]{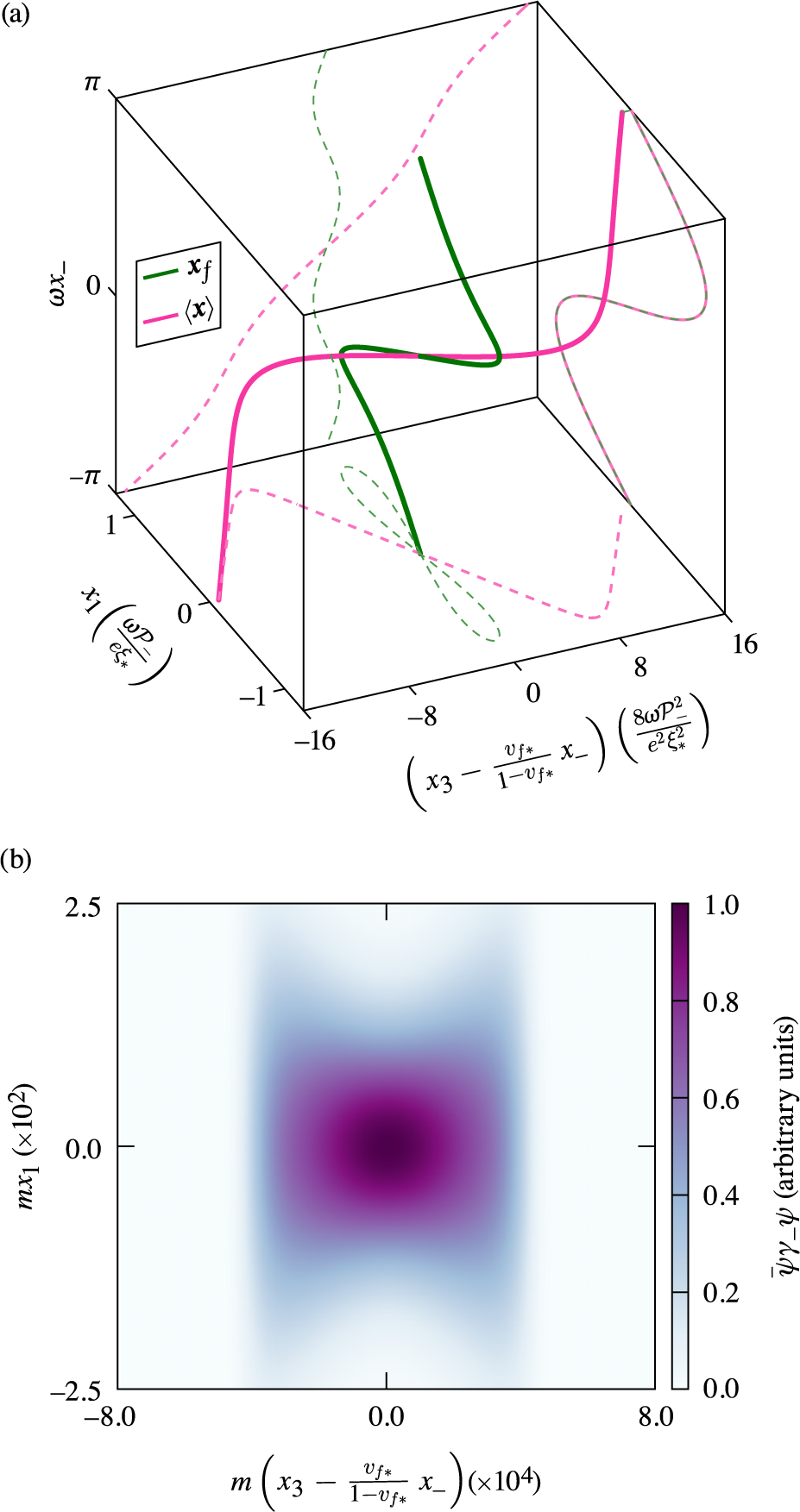}
\caption{Three-dimensional trajectory and probability density of the arbitrary-velocity Volkov wavepacket presented in Fig.~\ref{fig:2}(c) with $v_{f*}=-4.1$. The electromagnetic plane wave has a constant strength $ |e|\xi_*/m = 3$ and $\omega =0.01 m$. (a) The peak-probability trajectory (green) and expectation trajectory (pink) in a co-moving frame defined by $v_{f*}$ over one cycle $\omega x_- \in [-\pi,\, \pi]$. In this frame, the peak trajectory traces a figure eight, whereas in the co-moving frame defined by $v_{1\mathrm{D}}$, the expectation trajectory traces a figure eight (not shown), consistent with the classic result. (b) The probability density at $\omega x_- =0$. The transverse profile remains Gaussian while the maximum traces the figure eight. The minimum transverse width of the wavepacket is $w = 170 m^{-1}$ and ${\mathcal{P}}_- = 3m$.} \label{fig:3}
\end{figure}

Figure~\ref{fig:3}(a) compares the three-dimensional trajectories for the case presented in Fig.~\ref{fig:2}(c) over one cycle $\omega x_- \in [-\pi,\, \pi]$ near the maximum potential strength $ \xi = \xi_* $. In the frame defined by the target velocity $ {v}_{f*} = -4.1 $, the peak trajectory (green) traces a figure eight (projection on the bottom wall). This contrasts with the expectation trajectory, which exhibits more complex behavior, diverging from the peak both before and after their coincidence at $\omega x_- = 0$. If instead the frame were defined by the drift velocity $v_{1\mathrm{D}}$ [Eq.~\eqref{eq:lighfvel}], the expectation trajectory would trace out a figure eight closely resembling the classic result, whereas the peak would follow the more complex trajectory.

Figure~\ref{fig:3}(b) presents the numerically computed probability density at $\omega x_- = 0$. For large values of $w$, the density maintains a transverse Gaussian profile while its maximum traces the figure eight. For smaller values of $w$, the profile is nearly Gaussian at the beginning of each cycle, but becomes bowed along the trajectory due to terms of $\mathcal{O}(p_1 \mathrm{p}_\perp^2)$ in $S$ [Eq.\eqref{eq:FieldAction}]. These regimes are distinguished by the condition $w^2 \gg |\mathcal{P}_-^{-1}(1-v_a)/(\mathcal{P}_3 - v_a \mathcal{E})|$. Regardless of the $w$ value, Eqs.~\eqref{eq:xf}--\eqref{eq:zf} accurately predict the location of the peak.

%ABSOLUTLY SMALL KEEP FOR REF!
%W REQUIREMENT FOR SELF SIMILAR GAUSSIAN SHAPE ACROSS FIG 8
%\begin{equation}
%    w \gg \frac{e\xi_*}{\omega\mathcal{P}_-^{2/3}}\bigg| \frac{(1-v_a)}{(\mathcal{P}_3-v_a %\mathcal{E})}\bigg|^{1/3}
%\end{equation}
%RELATIVELY SMALL WIP
%\begin{equation}
%    \mathrm{p}_\perp^2 \ll 2 |(\mathcal{P}_3 - v_a %\mathcal{E})\mathcal{P}_-/(1-v_a)|
%\end{equation}

\begin{figure*}[htb]
\centering
\includegraphics[width=0.99\textwidth]{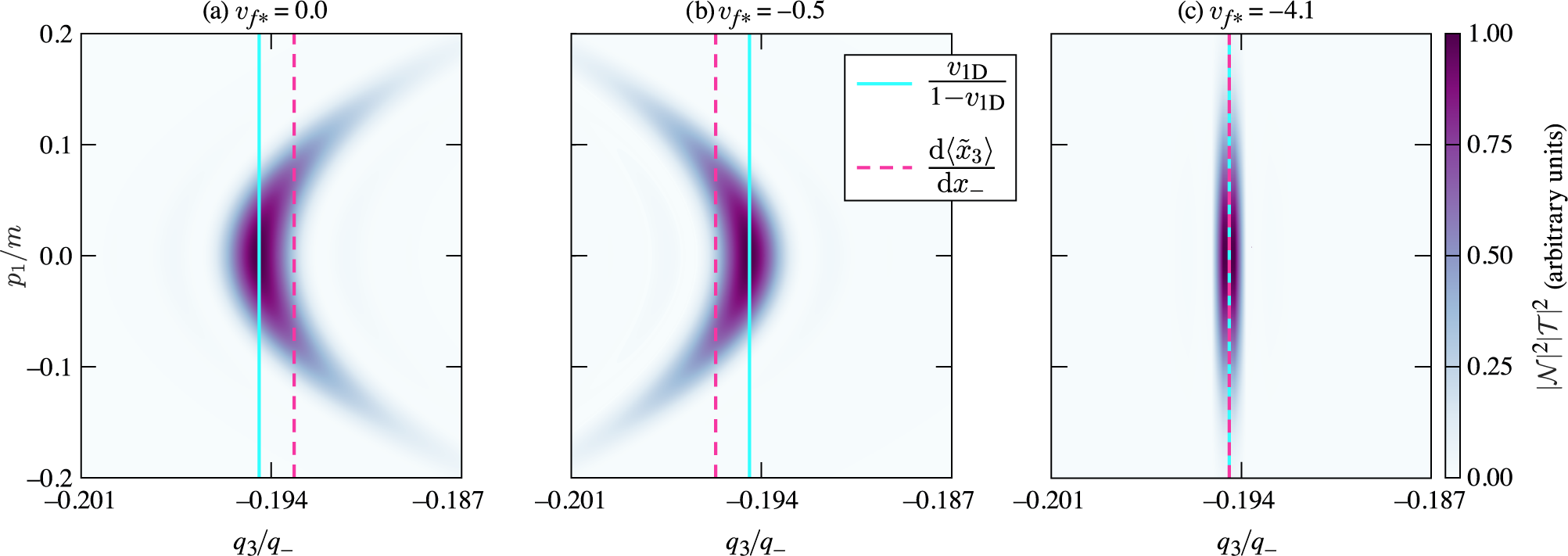}
\caption{The momentum density $|\mathcal{N(\eta)}|^2|\mathcal{T}(\mathbf{p}_\perp)|^2$ as a function of ${q}_3/q_-$ and $p_1$, illustrating the structure of the correlation built into the wavepacket for different values of $v_{f*}$. In (a), (b), and (c), the respective values $v_{f*} =0.0, -0.5,$ and $-4.1$ were chosen to produce distinct curvatures with respect to ${p}_1$. (a) The convex curvature shifts the expectation velocity (dashed pink line) above its one-dimensional value (solid blue line), $v_\mathrm{1D}/(1-v_\mathrm{1D})=-0.1944$ [see second term on the right-hand side of Eq.~\eqref{eq:lighfvel}]. (b) The concave curvature shifts the expectation velocity below its one-dimensional value. (c) The curvature vanishes in the special case $v_{f*} = v_{\mathrm{1D}}^{-1} =  -4.1$. Here, $|e|\xi_*/m = 3$, $v_\mathrm{1D}=-0.24$, ${\mathcal{P}}_- = 3m$, and $w = 4\pi\, m^{-1}$.  }\label{fig:4}
\end{figure*}

As discussed in the context of Eq.~\eqref{eq:lighfvel}, the expectation trajectory depends on the value of $v_{f*}$ and, by extension, $v_a$. This is a direct result of constructing the wavepacket with an out-of-field correlation between $p_3$ and $\mathrm{p}_\perp$ [Eq.~\eqref{eq:STcritp}] to produce an in-field correlation between $q_3$ and $\mathrm{p}_\perp$ [Eq.~\eqref{eq:STcritq}]. Figure~\ref{fig:4} shows that this correlation manifests as curved contours in $|\mathcal{N(\eta)}|^2 |\mathcal{T}(\mathbf{p}_\perp)|^2$, with curvature governed by $v_{f*}$. The values $v_{f*} = 0.0, -0.5,$ and $-4.1$ were chosen to illustrate distinct regimes. When the second term in Eq.~\eqref{eq:lighfvel} is positive, the curvature is convex, shifting the expected light-front velocity above the 1D result [Fig.~\ref{fig:4}(a)]; when it is negative, the curvature is concave, shifting the light-front velocity below the 1D value [Fig.~\ref{fig:4}(b)]. For $v_{f*} = v_{\mathrm{1D}}^{-1}$, the curvature vanishes, and the light-front velocity coincides with the 1D result [Fig.~\ref{fig:4}(c)]. In each case, the expectation velocity was obtained numerically from the cycle average of Eq.~\eqref{eq:EXPz}. Because the influence of the correlation on the expectation velocity scales inversely with $w$ [right-hand-side of Eq.~\eqref{eq:lighfvel}], a smaller $w = 4\pi m^{-1}$ was used than in Figs. \ref{fig:2} and \ref{fig:3}. 

%This signature of $v_{f*}$ and $v_a$ manifests as a permanent imprint in configuration space, best understood through a ``ray-trace perspective'' where a distant spherical detector records lepton arrival times as a function of incidence angle. In the absence of a field, while conventional wavepackets ($\eta = E_{\mathbf{p}}$) yield angle-independent arrival times due to uniform ray speeds, engineering an arbitrary-velocity wavepacket with separability $\eta = E_\mathbf{p}-v_a p_3$ creates a unique angle–arrival-time correlation. Crucially, a similar space--time structure emerges naturally among the dressed momentum when a conventional wavepacket interacts with an electromagnetic field. Although the ``dressed'' separability ($U_\mathbf{q} - v_f q_3$ and $\mathbf{p}_\perp$) is strictly transitory, vanishing once the interaction concludes, each propagation time through the field would depend on the outbound ray angle. Consequently, measuring this correlation shifts from validating an engineered wavepacket to observing a fundamental, intrinsic signature of the field interaction itself.

\section{Discussion and Conclusion}\label{sec:conclude}
The probability-density peak of a charged-lepton wavepacket can move through the field of an electromagnetic plane wave at an arbitrary, prescribed velocity. This is achieved by constructing the wavepacket with a specific momentum correlation prior to its interaction with the field. Upon entering the field, the initial structure gives rise to correlations between the dressed momenta of the constituent Volkov states. These correlations leave an imprint on the expectation trajectory, providing a potentially measurable signature of the prescribed velocity.

In configuration space and outside of the field, arbitrary-velocity Volkov wavepackets have constant spherical or hyperbolic wavefronts with respect to the comoving coordinate $x_3 - v_a x_0$. A potential experimental realization of such a wavepacket could involve introducing a time-dependent curvature phase at a fixed value of $x_3$. This can be accomplished using the Kapitza--Dirac effect. More specifically, two counter-propagating light waves with time-varying amplitudes form a standing wave that acts as a ``lens,'' focusing different temporal slices of the wavepacket to different locations. This mechanism is directly analogous to optical ``flying-focus'' techniques \cite{simpson2022spatiotemporal, ramsey2023exact, palastro2024superluminal, almeida2025universal}. In this context, $v_a$ can be interpreted as the ``focal'' velocity.

The theory of arbitrary-velocity Volkov wavepackets was presented for linearly polarized electromagnetic plane waves, and the examples performed for wavepackets with longitudinal momentum anti-parallel to the plane-wave propagation direction. However, both the theory and examples generalize to arbitrary polarizations, relative orientations, and peak velocity directions. Any in-field velocity vector $\bvec{v}_{f*}$ can be realized for a general plane-wave four-potential $A(x_-)$. Outside the field, this generalization is achieved by setting $\eta = p_0 - \mathbf{v}_a \cdot \mathbf{p}$. Because the transverse dressed momenta equals the transverse momenta, the in-field and out-of-field transverse velocities are identical. As a result, a specific $\bvec{v}_{f*}$ can be obtained by following the procedure in Sec.~\ref{sec:construct} C with $\xi^2_*/2$ replaced by the cycle average of $-A\cdot A$ at the target strength. As before, the component of $\bvec{v}_a$ along the propagation direction of the plane wave is chosen to ensure that the target velocity is reached at $\xi_*$. 

More broadly, this work highlights that electromagnetic fields induce spatiotemporal structure in charged-lepton wavepackets. Even a conventional wavepacket, separable in energy and transverse momentum, acquires a correlation among its dressed momenta during its interaction with an electromagnetic wave. These correlations cause the probability-density peak to follow a trajectory that deviates from the expectation trajectory. Thus, the dynamics of structured wavepackets are more than a product of wavefunction engineering; they are fundamental to the interaction of charged particles with electromagnetic fields. In fact, the structure of a conventional wavepacket is only preserved during interactions with stationary potentials. These results point to space--time correlations as a general feature of wavepacket dynamics, with implications beyond the leptonic systems studied here. For instance, analogous correlations would arise in conventional electromagnetic waves propagating through dynamic and structured media, where the refractive index acts as the potential.

\medskip
\begin{acknowledgments}
The authors thank B. Barbosa, A. Di Piazza, M. S. Formanek, S. Ahrens, K. Churnetski, and E. Melcher for valuable discussions.

This report was prepared as an account of work sponsored by an agency of the United States Government. Neither the United States Government nor any agency thereof, nor any of their employees, makes any warranty, express or implied, or assumes any legal liability or responsibility for the accuracy, completeness, or usefulness of any information, apparatus, product, or process disclosed, or represents that its use would not infringe privately owned rights. Reference herein to any specific commercial product, process, or service by trade name, trademark, manufacturer, or otherwise does not necessarily constitute or imply its endorsement, recommendation, or favoring by the United States Government or any agency thereof. The views and opinions of authors expressed herein do not necessarily state or reflect those of the United States Government or any agency thereof.

This material is based upon work supported by the Department of Energy [National Nuclear Security Administration] University of Rochester ``National Inertial Confinement Fusion Program'' under Award Number DE-NA0004144 and the Department of Energy Office of Science under Award Number DE-SC0021057. 
\end{acknowledgments}
%TC:endignore

\appendix
%\section{Velocity Limits of Eq.~\eqref{eq:STcritp}}
\section{Velocity Limits of Eq.~(14)}
This appendix examines three special cases of the velocity parameter: $v_a= \pm 1$ and $v_a = \mathcal{P}_3/\mathcal{E}$, with the latter corresponding to the singularity in Eqs.~\eqref{eq:parxExpandS} and \eqref{eq:lighfvel}. For $v_a = \pm1$, $\eta = p_{\mp}$ and Eq.~\eqref{eq:STcritp} has one finite branch \cite{palastro2024superluminal}
\begin{equation}
    p_3(\eta,\mathrm{p}_\perp) = \mp \frac{\eta^2 - m^2 - \mathrm{p}_\perp^2} {2\eta},
\end{equation}
which is real for all values of $p_\mp$ and $\mathrm{p}_\perp$. However, care must still be taken to exclude counter-propagating longitudinal content in the transverse distribution $\mathcal{T}(\mathbf{p}_\perp)$ by enforcing $\mathrm{p}_\perp^2 < \eta^2 - m^2$. The special case of wavepackets in an electromagnetic plane wave with $v_a = 1$ and $\eta = p_-$ was investigated in Ref. \cite{di2024radiation}.

For $v_a = \mathcal{P}_3/\mathcal{E}$, the on-shell longitudinal momentum evaluated at $\eta = \langle \eta \rangle$ is given by
\begin{equation}
    p_3(\langle{\eta} \rangle,  \mathrm{p}_\perp) =  \mathcal{P}_3 \pm i \frac{  \mathrm{p}_\perp \mathcal{E} }{m}.
\end{equation}
This marks the threshold at which $p_3$ acquires an imaginary component for any non-zero $\mathrm{p}_\perp$. To exclude these evanescent modes, the product $\mathcal{N}(\langle \eta \rangle)f(p,\langle \eta \rangle$) must be zero for all $\mathrm{p}_\perp > 0$. This can be accomplished in two ways: either $f(p,\eta) \propto \delta(\mathbf{p}_\perp)$ or $\mathcal{N}(\eta)$ vanishes at $\eta = \langle \eta \rangle$ with local maxima about this value. In the latter case, the expansions performed in the main text must be taken about each maxima to accurately characterize the spatiotemporal evolution of the wavepacket.

\section{Lifetime of the Peak-Probability Trajectory}
The trajectory of the probability-density peak is only apparent while it remains within the wavepacket envelope, which follows the expectation-value trajectory. Consequently, the length of the envelope, $\Delta x_3$, determines the lifetime of the peak. The length is inversely proportional to the spectral width $\Delta\eta$:
\begin{equation}
    \Delta x_3 \propto  \bigg| v_a-\frac{{\mathcal{P}}_3}{{\mathcal{E}}} \bigg|\frac{1}{\Delta\eta}.
\end{equation}
To leading order, the rising and falling edges of the envelope follow the trajectories
\begin{equation}
\begin{split}
    {\langle \tilde{x}_3 \rangle}_\mathrm{edge}\approx \frac{{\mathcal{P}}_3}{\mathcal{P}_-}x_- +\frac{e^2}{2 {\mathcal{P}}_-^2}\int_0^{x_-}\tilde{A^2_1}(x_-') \; \mathrm{d}x_-' \pm \frac{\Delta x_3}{1-{\mathcal{P}}_3/{\mathcal{E}}},
    \end{split}
\end{equation}
while the probability-density peak follows [Eq.~\eqref{eq:lightframefocal}]
\begin{equation}
    \tilde{x}_{f3} = \int_0^{x_-} \frac{v_f}{1-v_f} \; \mathrm{d}x_-'.
\end{equation}
The intersections of these trajectories, 
\begin{equation}
    \tilde{x}_{f3} - {\langle \tilde{x}_3 \rangle}_\mathrm{edge} = 0,
\end{equation}
determine the duration $\Delta x_-$ over which the peak persists. For a varying potential strength, these intercepts must be computed numerically. For a constant potential strength $\xi = \xi_*$, the duration can be computed analytically:
\begin{equation}\label{eq:dur}
\begin{split}
\Delta x_0 &= \frac{2\Delta x_3}{|1 - {\mathcal{P}}_3 / \mathcal{E}|} \left|  \frac{1 - v_{\mathrm{1D}}}{v_{f*} - v_{\mathrm{1D}}}  \right|,\\
\end{split}
\end{equation}
where $v_{\mathrm{1D}} $ is evaluated at $\xi= \xi_*$ [Eq.~\eqref{eq:v1D}]. In the absence of an electromagnetic field, $\xi_* = 0$, $ v_{\mathrm{1D}} = {\mathcal{P}}_3 / {\mathcal{E}}$, $v_{f*}=v_a$, and Eq. \eqref{eq:dur} reduces to
\begin{equation}
\Delta x_0 =  \frac{2\Delta x_3 }{|v_a- v_{\mathrm{1D}}|},
\end{equation}
consistent with the out-of-field result in Ref.~\cite{palastro2024superluminal}.

%% KEEP FOR REF
%\begin{equation}
%\begin{split}
%    \Delta x_3 &\propto \Delta \eta^{-1} \Bigg|\frac{\mathrm{d} \eta}{\mathrm{d}p_3}\Bigg|_{(\eta = \hat{\eta},\bvec{p}_\perp = \bvec{0})}\\
%    &=\Delta \eta^{-1}\Bigg|\frac{\hat{\mathcal{P}}_3}{\hat{\mathcal{E}}} -v_a \Bigg|,
%\end{split}
%\end{equation}

\clearpage
\bibliography{AVEVOS}
\bibliographystyle{apsrev4-1}
\end{document}